# Cybersecurity for Industrial Control Systems: A Survey

Deval Bhamare[Ω], Maede Zolanvari[φ], Aiman Erbad[Ұ], Raj Jain[φ], Khaled Khan[Ұ], Nader Meskin[£]

[Ω]Department of Mathematics and Computer Science, Karlstad University, Sweden
[Ұ]Department of Computer Science and Engineering, Qatar University, Doha, Qatar
[φ]Department of Computer Science and Engineering, Washington University in St. Louis, USA
[£]Department of Electrical Engineering, Qatar University, Doha, Qatar
deval.bhamare@kau.se, maede.zolanvari@wustl.edu, aerbad@qu.edu.qa, jain@wustl.edu, k.khan@qu.edu.qa, nader.meskin@qu.edu.qa

**Abstract:** Industrial Control System (ICS) is a general term that includes supervisory control & data acquisition (SCADA) systems, distributed control systems (DCS), and other control system configurations such as programmable logic controllers (PLC). ICSs are often found in the industrial sectors and critical infrastructures, such as nuclear and thermal plants, water treatment facilities, power generation, heavy industries, and distribution systems. Though ICSs were kept isolated from the Internet for so long, significant achievable business benefits are driving a convergence between ICSs and the Internet as well as information technology (IT) environments, such as cloud computing. As a result, ICSs have been exposed to the attack vectors used in the majority of cyber-attacks. However, ICS devices are inherently much less secure against such advanced attack scenarios. A compromise to ICS can lead to enormous physical damage and danger to human lives. In this work, we have a close look at the shift of the ICS from stand-alone systems to cloud-based environments. Then we discuss the major works, from industry and academia towards the development of the secure ICSs, especially applicability of the machine learning techniques for the ICS cyber-security. The work may help to address the challenges of securing industrial processes, particularly while migrating them to the cloud environments.

## I. INTRODUCTION

Industrial control systems (ICSs) are often found in the industrial sectors and critical infrastructures, such as nuclear and thermal plants, water treatment facilities, power generation, heavy industries, and distribution systems. Formally, ICS is a term that covers numerous control systems, including supervisory control and data acquisition (SCADA) systems, distributed control systems (DCS), and other control system configurations such as programmable logic controllers (PLC). An ICS is a combination of wireless and control components (e.g., electrical, mechanical, hydraulic, pneumatic), which achieve various industrial objectives (e.g., manufacturing, transportation of matter or energy). A typical ICS consists of numerous control loops, human-machine interfaces (HMIs), and remote diagnostics and maintenance tools built using an array of network protocols [1, 2]. From now on, we will be using the term "ICS" to represent all of its aforementioned components.

SCADA systems monitor and control different industrial control system components by collecting data from and issuing commands to geographically remote field stations. SCADA systems and DCS are often networked together to operate in tandem. Although the industrial facility operation is controlled by a DCS, the DCS must communicate with the SCADA system to coordinate production output with transmission and distribution demands (Fig. 1). For a long time, until very recently, ICSs were kept disconnected from the Internet. Communication among remote components happened over private networks and some specially designed protocols, for example, Modbus RTU, Modbus TCP, or various wireless technologies such as Wi-Fi, Z-Wave, Zigbee, and others. Other protocols also exist, such as common industrial protocol (CIP), Actuator-sensor interface (AC-i), DeviceNet, Highway Addressable Remote Transducer Protocol (HART) protocols for end-to-end automation. However, recently, industries are realizing the benefits that may be derived from the Internet as well as Information Technology (IT) environment, such as cloud computing [2].

Infrastructure as a service (IaaS) is a service offered by cloud providers, which is gaining significant attention recently. Industries can benefit from such cloud-based services. Instances of SCADA systems and PLC controllers can be implemented as a service using the infrastructure provided by the clouds, through IaaS. This might save significant hardware and infrastructure cost for the industries. However, connecting ICS components to the cloud and the Internet exposes ICSs to the majority of cyber-attack scenarios. Moreover, ICS devices are inherently less secured against such advanced attacks compared to the traditional ICS attacks (e.g., catastrophic human error or insider sabotage). This is due to their different characteristics and thread complexity.

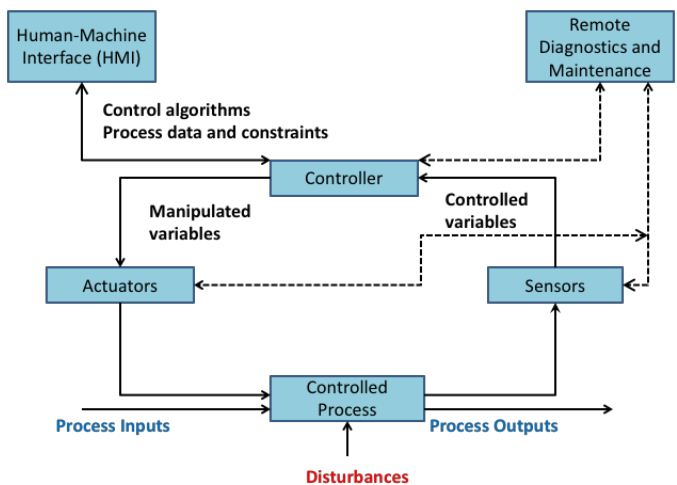

Fig. 1. Components of an ICS.

Examples of potential ICS-related threats include [61]:
- Advanced Persistent Threats (APTs)
- Unintended spillover of corporate network compromises
- Disruption of voice & data network services
- Coordinated physical & cyber-attack
- Hacktivist attacks
- Supply chain disruption or compromise
- Distributed Denial of Service (DDOS)

Moving towards cloud-computing for ICSs has several advantages, such as scalability, cost-efficiency, flexibility. However, while moving towards the cloud, ICSs may get exposed to new threats and vulnerabilities [88], [89].
- ICS managers have limited security controls over the data. Therefore, loss of data privacy and opening up an easy point of illegal access to the assets.
- Loss of connection with the remote components from the local devices or vice versa. Therefore, a threat of loss of data, delays in the production process, propagating error and negatively impacting other related sections, etc.
- Abusing the current flaws in the local security controls by other remote cloud users. Therefore, a threat of data breach, data theft, data manipulation, data exploitation, etc.
- Lack of security standardization for cloud-based ICSs. All the components must follow the same security framework to provide consistency in operation.

Existing defenses such as firewalls and VPNs have repeatedly proven inadequate on their own, especially with the increasing usage of cloud platforms for the ICSs. In general, data encryption technologies to secure data transmission over the network have been proposed. Various encryption algorithms have been adopted in the literature, such as Data Encryption Standards (DES), Triple DES, Advanced Encryption standards (AES), Blowfish, and others. A general scheme of the performance comparison for such algorithms is provided in [52]. Other advanced techniques such as digital signature have been proposed in the literature and used in cloud computing for additional security, which can be applied to the ICSs as well [53]. A significant amount of work has already been done in academia to build secure cloud-platforms for control systems. Secure and scalable access control for cloud computing has been proposed in [54]. Authors of [54] achieve this goal by exploiting and uniquely combining techniques of attribute-based encryption (ABE), proxy re-encryption, and lazy re-encryption. On the other hand, various research works have argued and demonstrated that data encryption alone is not sufficient for network security [55]. Recently there has been a trend in the applications of machine learning techniques in developing the intrusion detection systems (IDS) for ICS.

In this work, we focus on securing cloud platforms for ICSs to detect any malice or abnormal behavior. In other words, we study about building a secure ICS and advanced intrusion detection system (IDS) for ICS against the cyber-attacks. It is imperative that first, we have a closer look at the cloudification of the ICSs and recent works, which address the cyber-security issues in the ICSs. Our major focus in this work has been applicability of the machine learning techniques for anomaly and intrusion detection in the ICS and SCADA systems. In addition, we propose a simulation testbed to collect the data traces for ICS operations to build rigorous machine learning models.

The rest of the paper is organized as follows. In Section II, we critically analyze the survey work available in the field of ICS security. In Section III, we discuss the major research works in academia and industry on the application of cloud computing in ICSs, the cloud-related security issues, and the proposed solutions. Section IV lists the machine learning approaches for the cyber-security of the ICSs. In Section V, we present several case studies while in Section VI we argue for the need of a testbed to collect the data traces for ICS systems to aid building rigorous machine learning models. In Section VI, we also discuss future research directions. Finally, we conclude the paper. Table 1 summarizes the related work critically while the list of acronyms used in the paper is given in Table 2 at the end.

## II. AVAILABLE SURVEYS

In this section, we critically analyze the major surveys in the field of industrial control systems (ICSs) and their security and justify the need for further research. Kriaa et al. [84] have provided an extensive survey in the field of safety and security of industrial control systems. The borderline between these two concepts (safety and security for ICSs) has been clearly defined. Different approaches proposed in the literature for these issues are also categorized as either generic or non-generic. Cherdantseva et al. [77] review state of the art in cybersecurity risk assessment of Supervisory Control and Data Acquisition (SCADA) systems as well. However, both these works lack the survey of cloud-based approaches and machine learning techniques for ICS security, which is the main focus of research currently.

Knowles et al. [85] have surveyed the cyber-security of ICSs and the risk management aspects of it. The related standard in this domain are discussed and how the current systems lack built-in security considerations. Sajid et al. [78] have provided a review focusing on the security challenges of cloud-based ICS systems. Remarks such as additional challenges after cloud integration along with the general security shortcoming of SCADA systems are mentioned. However, a more in-depth security analysis is required, especially the applicability of the machine learning approaches.

Ding et al. [86] have provided a survey on the developed mathematical approaches for distributed filtering and control of ICSs. The main focus is on the differential dynamic models, while a small section is dedicated to security controls. There is a need to develop model-based approaches for the security of these systems. Molina and Jacob [87] have surveyed the current techniques based on software-defined networking solutions for the cyber-security of industrial settings. However, their focus is on the general concept of cyber-physical systems and not specific on ICSs. Zeng and Zhou [90] have studied the available approaches for intrusion detection systems (IDSs) deployed in ICSs. A taxonomy of the relevant vulnerabilities in these systems is also brought up. They discuss machine learning-based solutions along with other types of IDSs. However, none of the mentioned techniques are cloud-specific, and more detailed work regarding the available machine learning solutions for multi-cloud scenarios is required. In the next section we focus on the related work in the literature, more specific to multi-cloud deployment of the industrial control systems.

## III. RELATED WORK

Moving towards the cloud-based environments benefits both the control system providers and the users significantly by reducing the cost and increasing the throughput. In addition, higher reliability and enhanced functionality will be achieved by embracing cloud-based approaches for ICSs. In addition to eliminating the expenses and complexities associated with the hardware layer of the ICS infrastructure, a cloud-based control system allows connectivity among the remote components as well as enabling the users to utilize the system remotely [63]. SCADA, PLC and DCS systems, with their reliance on proprietary network protocols and equipment, have long been considered immune to the network attacks that have emerged recently in the networking paradigm. However, recent studies prove that this assumption is not correct.

Recently, standards such as Ethernet, TCP/IP, and web technologies have been compromised to attack the ICSs. Works presented in [2] and [40] demonstrate that there is a definite shift in the source of cyber-attacks on the ICSs. Threats originating from outside the organization are likely to have very different attack characteristics compared to the internal threats. Regarding the studies by the industry, such as IBM, attacks targeting the ICSs are already up by 110 percent [64]. Specifically, the spike in malicious ICS traffic was related to SCADA brute-force attacks, which uses automation to attack the system. Once the system is breached, attackers can remotely monitor or control the connected SCADA devices. Some recent notable ICS attacks include the cyber-attacks on the European Energy Company, New York dam attack, Russian cyber-attack on the Ukrainian power grid [64].

The associated costs of security breaches in real-time industrial platforms could be very high. They result in the loss of sensitive data, loss of revenues, environmental impacts, production, and financial loss, and even human injury. Hence, companies and industries are mandated to reassess their security risk models and their assumptions. It has become imperative to have a closer look at the state of the art of the security options for ICSs. This helps to better understand the current risks associated with the enhancements offered by the novel cloud technologies and the available solutions. In this section, we investigate the major works in academia and industry on the cloud computing applications in ICSs as well as the cloud-related security issues and available solutions. Such research works might encourage the industries to shift from the standard hardware systems to cloud-based platforms to leverage their advantages.

As mentioned before, there have been significant efforts in the industry as well as academia towards shifting the standard ICS system from the monolithic hardware to the cloud-based platforms. Pollet [9] has discussed the security strategies for SCADA in his work. The author describes specific threats to SCADA systems and some sample hacking scenarios and then ranks these threats in a security matrix. Igure et al. [7] highlight some major threats and vulnerabilities in the SCADA systems. The authors also provide a brief overview of some of the ongoing works in this field along with ongoing technical problems that should be addressed to improve the overall security of SCADA networks. Nicholson et al. [8] identify the security issues related to SCADA from the cyber-security perspectives. Attack mitigations, standards, and forums related to SCADA systems are also discussed.

Chandia et al. [10] propose two practical strategies for the security of the SCADA systems. The first strategy utilizes a security service suite that minimizes the impact on time-critical industrial processes while adhering to industry standards. The second strategy implements a sophisticated forensic system for collecting and analyzing the SCADA network traffic. Recent attacks on SCADA systems highlight the need for a SCADA security testbed, which can be used to model a real SCADA system and study the effects of attacks on them. Queiroz et al. [11] propose an architecture for a modular SCADA testbed and describe a tool, which mimics a SCADA network, monitors and controls the sensors and actuators using Modbus/TCP protocol. Bowen et al. [12] discuss the next-generation security issues in the SCADA systems and discuss best practices to avoid the attacks on the SCADA systems.

Wang et al. [14] confirm that simulation experiments are a potential means of analyzing and assessing the security of SCADA systems.

However, existing simulation environments have flexibility and extensibility limitations. In that work, the authors establish abstract models of a SCADA system based on the industry standards and propose a reference architecture of the SCADA simulation environment. Some experiments on attack scenarios have been conducted in the proposed simulation environment. Finally, they have analyzed and assessed the system's security status, and the results to demonstrate the effectiveness and practicability of the simulation environment [6]. Shahzad et al. [15] propose cryptography-based solutions to protect SCADA protocols and communications. Attack (abnormal) scenarios have been created within the testbed implementation (cryptography testbed), and the level of security is measured and compared based on the attack detection level and attack impact level. Qin et al. [16] design a neuron model, which combines reasoning machine based on the cloud generator with the factor neural network theory. The authors claim that the proposed system demonstrates substantial flexibility, mobility, scalability and other features across multiple platforms with distributed characteristics.

Zhang et al. [17] propose a fuzzy Petri network-based security defense model for SACADA systems. The authors also demonstrate the effectiveness of the proposed method with their results. The authors in [18] propose the need to extend the existing power-grid security models to secure the distributed smart grids. Colombo et al. [19] discuss the cloud-based cyber-physical ICSs. Similarly, the authors in [20] present a testbed platform for the implementation of cloud computing in SCADA systems for electrical power systems. The authors demonstrate the operations of the traditional SCADA systems over the cloud computing platforms and integrate the main application modules of SCADA for electrical power metering on the cloud platforms. Different types of control center servers along with physical networking solutions of the cloud platforms are also visualized. Similar work has been presented in [22] on implementing the SCADA systems over the cloud platforms. These are some of the pioneering works demonstrating the usability and advantages of the cloud platforms for the ICSs. Simmhan et al. [21] propose a cloud-based software system for big data analytics in the smart grids. The authors highlight the feasibility and effectiveness of such systems.

There are several works that investigate the security issues of cloud-based SCADA systems. For example, authors in [23] propose a secure cloud-based platform for SCADA systems considering the use case of the water supply networks. The authors present a proof of concept for a cloud-based water supply network monitoring (WSNM) application, called RiskBuster (RB). The proposed platform ensures the confidentiality and integrity of the SCADA system while monitoring and collecting the data from dam sensors and storing in the cloud by using the innovative Intel Software Guard eXtension (SGX) technology. Sajid et al. [24] highlight the security challenges in the cloud-based SCADA systems. Authors also provide the existing best practices and recommendations for improving and maintaining security along with future directions to secure the cyber-physical systems.

Shahzad et al. [25] have implemented a SCADA system within a cloud computing environment to minimize the costs related to real-time infrastructure for SCADA implementation through cloud computing. Cardenas et al. [62] have performed a risk assessment for SCADA systems and discussed various attack models with experiments. Authors have categorized the attacks in various groups and discussed the response schemes to such attacks. The authors in [26] identify the service-oriented architecture paradigm empowered by resource virtualization as a lighthouse for cloud-based ICSs. The proposed framework has the potential of empowering the seamless integration and interaction among the heterogeneous stakeholders in the future industrial automation domain. They have fulfilled this concept by integrating web services, Internet technologies, cloud systems, and the power of the Internet of Things. A similar security-oriented cloud-based SOA platform for ICSs has been proposed by Baker et al. in [27]. Qui and Gooi [42] propose a web-based SCADA display system through which users can view as well as control the operations of the sub-stations at the server sides. The authors in [2] identify various sources of possible threats to the ICSs, i.e., adversarial, accidental, structural, and environmental. Such source categorizing helps to create a risk management strategy that protects the system against possible threats.

Recently, there has been a trend towards implementing machine learning techniques for anomaly detection and prevention in the networks [5]. Karnouskos et al. [28] discuss trends and challenges for cloud-based industrial cyber-physical systems. The authors identify machine learning as a key trend in the security implementation for the cloud-based ICSs. It has been shown that machine learning methods are indeed able to provide useful security information for various physical problems and practical contexts. Thames and Schaefer [79] have investigated the integration of software-defined networks (SDN) in ICSs. Their proposed cloud-based workflow is called Software-Defined Cloud Manufacturing (SDCM). Utilizing the abstraction, provided through SDN architecture, provides low latency and efficiency in updating the security solutions and hardware designs. All these outcomes are promising in improving the manufacturing processes.

There has been a great effort in identifying the threats and attack vectors specific to ICSs. Rubio et al. [80] first discuss the common threats exposed to industrial control systems, and then the corresponding potential defense mechanisms are highlighted. In this paper, a threat analysis specific to cloud-based computing in these systems has been conducted. Zhou et al. [81] have studied one of the most relevant attacks against these systems, DDoS. They have utilized a fog-based computation to run real-time traffic monitoring and low latency communication with the cloud center for proper mitigations. As per the studies in [5], authors argue that automatic machine learning approaches are more systematic, easier to handle and master, and therefore, more reliable and robust than the traditional security measures such as firewalls. These possibilities open up new perspectives to respond to the challenges of planning and operating future industrial systems with an acceptable level of security. In the next section, we will have a closer look at the security measures for the ICSs using machine learning approaches.

## IV. MACHINE LEARNING APPROACHES FOR ICS SECURITY

With the discovery of the Stuxnet attack, increasing attention is being paid to the potential malware targeting the PLCs. Contemporary advanced malware may infer the structure of the physical plant and can use this information to construct a dynamic payload to achieve an adversary's end goal, such as Stuxnet [56, 57]. To counter such situations, recently, there has been a trend towards the implementation of machine learning techniques for anomaly detection and prevention in the networks of the ICSs [58-60].

ICSs, such as SCADA systems, have very regular communication patterns. Often the same limited set of read and write commands are repeated in a loop. For example, the gas pipeline system presented by Zhang et al. [17] repeats the same two commands in a loop. First, it writes the contents of all registers and coils used for control. Next, a Modbus read command is used to read the measured state of the system. Modbus protocol is often used in many ICSs, specifically SCADA systems. It implements a master/slave configuration in which the slave does not request the data. It only receives commands from the master. Data is transmitted over serial lines (Modbus RTU/ASCII) or Ethernet (Modbus TCP). These two commands are followed by a set of two responses. Such regularity leads to a set of commands in which all device addresses are constant, and each of the four packets has the same length. This lack of variation is expected. These regular patterns can be exploited by machine learning algorithms, which can be used to build models of normal behavior and detect abnormal deviations [1].

As per the studies in [5], it has been observed that machine learning methods are indeed able to provide important security information for various physical problems and practical contexts. Authors argue that automatic machine learning approaches are more systematic, easier to handle, and master. These possibilities open up new perspectives to respond to the challenge of planning and operating future industrial systems with an acceptable level of security. Brundle and Naedele [3] discuss the importance of securing the ICSs by listing the challenges of SCADA security. Recently, researchers have started studying intrusion detection systems using SCADA network traffic traces. With the advent of the novel multi-cloud networking paradigm, subsets of ICSs may be shifted to the clouds leveraging the advantages of the IaaS platform offered by various cloud service providers [63, 67]. However, security issues in such novel platforms need to be addressed before this major shift. To achieve this, more attention must be paid to studying machine learning applications for anomaly detection in multi-cloud environments.

Contemporary IDSs use machine learning algorithms for pattern recognition to detect threat activities that are anomalous for a particular system. There are other IDSs, which use signature-based systems to compare the activities to a database of known threats [29-31]. These functionalities can be combined together for a robust detection system and will provide a sufficient layer of protection for various attack scenarios. Despite the popularity of the machine learning techniques, IDS researcher groups lack standard datasets to train and test the algorithms. This results in an inability to develop robust ML models to detect anomalies in the ICSs [1]. Many of the datasets, especially in the context of the ICSs, used by researchers do not contain all types of attacks, hence gauging the performance of the IDS is hard when all patterns of attack are not considered, for instance, work presented in [4].

The authors in [1] develop a standard dataset to provide third-party validation of the IDS solutions. The dataset created from this research was purposed to fill the void in this area. However, a significant amount of future research is still required in this domain. Authors in [29] propose an IDS for SCADA protection based on machine learning approaches. Specifically, the authors compare rule-based approaches, ANN (Artificial Neural Networks), HMM (Hidden Markov Model) and SVM (Support Vector Machines). Similarly, authors in [30] propose an OCSVM (One-Class SVM) intrusion detection mechanism, which is based on unsupervised machine learning methods, that does not need any labeled data for training or any information about the anomaly. Dua and Du [31] propose data mining-based machine learning techniques for cybersecurity in SCADA systems.

In [1], the authors argue that IDS researchers lack a common framework to train and test the proposed algorithms. The authors try to bridge the gap by documenting two approaches of data sharing for the IDSs of the ICS research community. In their work, first, a network traffic data log captured from a gas pipeline is presented. The log was captured in a laboratory and included artifacts of normal operation and cyber-attacks. Second, an expandable virtual gas pipeline is presented which includes an HMI, PLC, Modbus/TCP communication, and a Simulink based gas pipeline model. The virtual gas pipeline provides the ability to model the cyber-attacks and normal behavior. In total 35 cyber-attacks were used in the creation of the data log. Also, the data log contains records from 214,580 Modbus network packets with 60,048 packets associated with cyber-attacks. In the data logs, each packet is labeled with the attack number or the label 0 for packets associated with normal events (no attack) [1]. Different models using the simulation tools can be prepared to represent the traffic in the related ICSs, which can then be used to train the machine learning models.

Hink et al. [33] have demonstrated the use of machine learning techniques for power system disturbance and cyber-attack discrimination. The authors in [32] focus on the detection of cyber-attacks in water distribution systems using machine learning techniques. A simple one-class classification approach in the feature space is proposed. The tests are conducted on a real dataset from the primary water distribution system in France, and the proposed approach is compared with other well-known one-class approaches. Similarly, authors in [34] propose the use of machine learning techniques by considering the diagnosis of wind turbine faults. The authors have demonstrated the use of classification techniques to predict the faults in advance. Beaver et al. [35] compare and evaluate various machine learning algorithms for anomaly detection in SCADA communication channel. Erez and Wool [36] describe a novel domain-aware anomaly detection system that detects irregular changes in Modbus/TCP SCADA control register values. The research focuses on discovering the presence of three classes of registers (sensor registers, counter registers, and constant registers) using the proposed automatic classifier. Additionally, parameterized behavior models are developed for each class.

Zhang et al. [37] propose a SCADA intrusion detection system based on self-learning semi-supervised OCSVM (S2 OCSVM). The authors demonstrate that S2 OCSVM can improve detection accuracy. The authors in [38] propose anomaly detection in electrical substation circuits via unsupervised machine learning methods. The authors present preliminary results of characterizing normal, faulty, and attack states in smart distribution substations. However, further investigation for the broader applicability of the proposed methods is needed. The authors in [39] propose an architecture of an anomaly detection system based on the Hidden Markov Model (HMM) algorithm for intrusion detection in ICSs, especially in SCADA systems interconnected using TCP/IP.

Table 1 summarizes the research works performed in the context of cloud-based industrial systems and approaches for the cyber-security of the ICSs. Besides academia, various industries have taken the initiative for the security of the ICSs. In the remainder of this section, we describe some of the important industry efforts in the field of cybersecurity for the ICSs. Cisco Systems Inc. has developed an open-source Linux-based firewall that is capable of filtering Modbus packets. The firewall adds Modbus functionality to Linux's Netfilter tool that addresses the SCADA protocol. Authors at Cisco have proposed to detect malicious behavior based on the analysis of network proxy logs using weak supervision methods [47]. In another paper, [48], the same authors claim that weak supervision can be adopted on the level of properly defined bags of proxy logs by leveraging the Internet domain blacklists, security reports, and sandboxing analysis. The authors focus on training the detectors to identify the malicious URLs in the network traffic by using the proxy logs of HTTP requests. They have utilized a classification system that uses statistical feature representation computed from the network traffic and learn to recognize malicious behavior.

Table 1: Summary of approaches for ICS/SCADA cyber-security

| Work | Description |
|---|---|
| **Category**: Implementation of Machine learning Algorithms, Academia | |
| [5] | Use of decision tree induction, multilayer perceptron and nearest neighbor classifiers for power system security assessment |
| [62] | Attack categorization and IDS for SCADA security. |
| [31] | Data mining-based machine learning techniques for cybersecurity |
| [29] | IDS using ANN, HMM and SVM |
| [35] | Compare and evaluate various machine learning algorithms for anomaly detection in SCADA communication channel |
| [30] | IDS using One-Class Support Vector Machine |
| [33] | Demonstrate the use of machine learning techniques for power system disturbance and cyber-attack discrimination |
| [37] | Self-learning semi-supervised one-class Support Vector Machine (S2 OCSVM) system for intrusion detection in SCADA |
| [36] | Domain-aware anomaly detection system to detect irregular changes in Modbus/TCP SCADA control register values |
| [47] | Weak supervision techniques on the network logs for network traffic security |
| [32] | Detection of cyber-attacks in water distribution systems using machine learning techniques |
| [34] | Use of classification machine learning techniques for the diagnosis of wind turbine faults in advance. |
| [38] | Anomaly detection in electrical substation circuits via unsupervised machine learning methods. |

| Ref | Description |
|---|---|
| [39] | The architecture of an anomaly detection system based on the Hidden Markov Model (HMM) for intrusion detection in ICS |
| [48] | A classification system that uses the statistical feature based on SVM |
| **Category**: SCADA system architecture, protocols, and security, Academia | |
| [8], [9] | Identification of the security-related issues for SCADA from the cyber-security perspective |
| [42] | A web-based SCADA display system is proposed through which users can view as well as control the operations of the sub-stations at the server sides |
| [12] | Next-generation security issues in the SCADA systems are discussed |
| [10] | Two practical strategies for the security of the SCADA systems |
| [11] | The architecture of a modular SCADA testbed and a tool, which mimics a SCADA network, monitors and controls real sensors and actuators using Modbus/TCP protocol has been described |
| [13] | The PowerCyber testbed has been designed to resemble power grid communication utilizing actual field devices and SCADA software |
| [14] | A reference architecture of SCADA system simulation environment |
| [21] | A cloud-based software system for the big data analytics in the smart grids |
| [15] | Cryptography based solutions to protect the SCADA protocols and communications |
| [19] | A cloud-based cyber-physical industrial control system |
| [25] | Implemented a SCADA system within a cloud computing environment, to minimize the cost that is related to real-time infrastructure or SCADA implementation |
| [26] | Identify the service-oriented architecture paradigm empowered by virtualization of resources as a lighthouse for cloud-based ICSs |
| [28] | Discuss trends and challenges for Cloud-based industrial cyber-physical systems. Also, identify machine learning as a key trend in the security implementation for the cloud-based industrial control systems |
| [27] | A security-oriented cloud-based SOA platform for ICSs has been proposed. |
| [17] | A fuzzy Petri network-based security defense model for SACADA systems |
| [24] | Best practices and recommendations for improving and maintaining security along with future directions to secure the cyber-physical systems are discussed |
| [23] | Use of Intel Software Guard eXtension (SGX) technology for confidentiality and integrity of SCADA monitoring data. |
| [20] | A testbed platform for the implementation of Cloud Computing in SCADA systems for Electrical Power Systems |
| **Category**: Implementation and Standardization, Industry | |
| [43] | Information security and risk assessment for the ICSs by PSCC |
| [44] | 21 steps by Department of Energy, USA, to make the SCADA systems secure |
| [45] | Standards for managing the data access, alarms, event management, and even Web access to the SCADA network devices by OPC foundation |
| [47] | An open-source Linux-based firewall that is capable of filtering Modbus packets by CISCO |
| [48] | Training the detectors to identify the malicious URLs in the network traffic by using the proxy logs of HTTP requests by CISCO |
| **Category**: Dataset generation to train machine learning algorithms, Academia | |
| [73] | Lawrence Berkeley National Lab (LBNL) dataset |
| [68] | ISOT dataset |
| [76] | Information Security Center of Excellence (ISCX) |
| [74] | UNSW (University of New South Wales) dataset |
| [1] | Setup for dataset generation |
| [75] | next-generation IDS dataset (NGIDS-DS) |
| **Category**: IEEE Standardization | |
| [49] | IEEE standard for security systems for nuclear power generating stations |
| [50] | IEEE Guide for electric power substation physical and electronic security |
| [65] | IEEE guide for Security aspects regarding the access, operation, configuration, and data retrieval from IEDs |

Department of Energy, USA, has identified 21 steps to make the SCADA systems secure [44]. The OPC Foundation is also another organization working towards the development of security standards for regulating the client-server access and open connectivity in industrial automation. OPC Foundation started as OLE (object linking and embedding) for Process Control. It has developed standards for managing data access, alarms, event management, and even Web access to the SCADA network devices [45].

IEEE Power Engineering Society (PES) Power System Communications Committee (PSCC) has also done significant work in the domain of information security and risk assessment for the ICSs [43]. An instance is the IEEE standard for security systems for nuclear power generating stations [49]. This standard addresses the equipment for security-related detection, surveillance, access control, communication, data acquisition, and threat assessment. Security issues related to the human intrusion upon electric power supply substations are identified and discussed in E7.1402 - the IEEE Guide for electric power substation physical and electronic security [50]. Security aspects regarding access, operation, configuration, firmware revision and data retrieval from intelligent electronic devices (IEDs) are addressed in IEEE 1686 [65].

IEEE Power Engineering Power System Communications and Cybersecurity (PE/PSCC) Working Group C1 has developed a standard for cybersecurity requirements in control systems [51]. Cybersecurity measures require that a balance be achieved between the technical feasibility and the economic feasibility and that this balance addresses the risks expected to be present at a substation. Further, cybersecurity measures must be designed and implemented in such a manner that access and operation to legitimate activities are not impeded, particularly during times of emergency or restoration activity. IEEE presents a balance of the above factors in this standard.

ISA/IEC standardization bodies with the ISA99/IEC 62443 have taken an initiative towards the security of ICSs even though some industries in few countries have taken more focused initiatives, such as API (American Petrol Institute), AGA (American Gas Association) or IAEA (International Atomic Energy Agency) for nuclear technology [45]. France has even developed protection profiles (a concept drawn from ISO 15408 evaluation criteria for IT security) for ICS components. Numerous security solutions such as data diodes and industrial firewalls exist, but they usually address the upper part of an ICS classical architecture (ISA 95 model). Security solutions, in the context of communications between the PLCs and the actuators/sensors, installed by an exploit (in particular, if they are related to internal threats) have yet to be developed and deployed.

Security products providers such as FireEye, Darktrace, and Cisco have included machine learning in their security-related products and services [61, 66]. Similarly, other organizations such as North American Electric Reliability Corporation (NERC), International Organization for Standardization (ISO), International Society of Automation (ISA) have also contributed towards the development of the standards for the cybersecurity of the SCADA and PLC systems. In the next section, we discuss four different case studies presented in four different research works available in the literature.

## V. CASE STUDY EXAMPLES

In this section we discuss four specific case studies for multi-cloud deployment of ICSs and applicability of machine learning techniques for ICS security, which are: (1) Cyber risk assessments based on machine learning for the ICSs [82], (2) Cloud-based computing in these systems and how it improves the attack mitigation [81], (3) Comparison of several machine learning methods to detect malicious SCADA communications [35] and (4) Thorough step by step cloud mitigation of the SCADA systems [91].

The relevant threats in ICSs have been investigated comprehensively in [82] with a case study. Challenges in the way of utilizing machine learning and how it can help in defense mechanisms are also discussed. Nine different prevalent attacks along with reasons whether machine-learning-based approaches are useful or not are discussed. One of the main obstacles in utilizing a learning model for real-world industrial setups is imbalanced datasets [83]. An imbalanced dataset simply means the number of instances in one class is significantly lower than the other class. If the machine learning model does not train with enough samples from a particular class, it fails in real-time detection of that particular class, and it would misclassify them. This is a bottleneck in ICSs since most of the time, the number of attack samples is relatively very low compared to the massive amount of normal traffic flowing in the network. Utilizing common and traditional machine learning models will result in a large number of false negatives (attack traffic classified as normal). Based on the results in [82], machine learning is useful when the traffic flow is somehow manipulated. For instance, in confidentiality attacks, where the intruder just eavesdrops on the network traffic, machine learning might not be the best tool to detect that sort of breach.

A cloud-based framework for ICSs for DDoS mitigation is discussed in [81]. Mitigation simply means the attack is discovered, and then appropriate countermeasures are applied. The authors have worked on a case study of integrating fog computing into cloud computing for faster results. A testbed has been simulated to demonstrate the performance. All parts of an ICS (the field devices, e.g., sensors and actuators, RTUs) are simulated. Fog environment and cloud environment using a cloud server are also simulated to run the experiments. The effectiveness of the scheme is tested against two different DDoS attacks. A flooding technique, TCP SYN, and exploitation of the Modbus's vulnerability with forged command data are the two types of deployed DDoS attacks. The proposed scheme shows an improvement of average 8.98% in the detection rate, demonstrating a good improvement in the performance.

Six Different types of machine learning approaches for IDS of SCADA systems are evaluated in [35]. These techniques include Naive Bayes, random forests (RF), OneR, J48, NNge (Non-Nested Generalized Exemplars), SVM (support vector machines). In this specific case study, the authors have used labeled RTU telemetry data from a gas pipeline system in Mississippi State University's Critical Infrastructure Protection Center. Attack traffic is simulated to test the machine learning performance and is generated from two types of code injection sets including command injection attacks, and data injection attacks. Seven different variants of data injection attacks were tried to change the pipeline pressure values, and four different variants of command injection attacks to manipulate the commands that control the gas pipeline. Two performance metrics, precision and recall have been used for evaluation. The imbalanced ratio of the proposed dataset is about 17%.

The performance of the machine learning techniques is evaluated for the two types of attacks separately. The results of their case study show that for data injection attacks, NNge, and random forest have the highest score for both metrics. However, for the command injection, all the models show consistent and almost the same performance. The authors have declared that this is due to the simplicity of the command injection attacks compared to data injection attacks.

In the fourth case study [91], the authors have done an excellent analysis by presenting how a SCADA system can be deployed over an Infrastructure as a Service (IaaS) cloud setting. They have utilized EclipseSCADA as the open-source base system and the cloud server providing the IaaS is NeCTAR. Comprehensive comparison among different available open-source SCADA platforms has also been conducted. They have compared different aspects such as architecture and feature, code manageability, and adoption. Therefore, EclipseSCADA was chosen as it provides enough flexibility for cloud infrastructure. This platform and its components have been explained with details. However, the field devices (sensors and actuators) have been simulated.

In the evaluation setup, a complete set of performance metrics are investigated. Since in an ICS environment, real-time monitoring is crucial, they have used metrics such as the effect of the size of the monitored field and delays, the effect of the location of SCADA components, and the relationship between the processing time and the communication time. Various experimental settings were tested out to provide a thorough recommendation list for moving to the cloud in a real-world scenario. For instance, centralizing all the functionalities is not useful even though it sounds promising. Moreover, to lower the network load, event-driven communication protocols must be utilized. It is suggested to do the protocol conversion close to the field devices, and a lot more that can be further studied in the paper. In the next section, we propose a simulation testbed to collect the data traces for industrial control system operations to build rigorous machine learning models and discuss future research directions.

## VI. RESEARCH CHALLENGES

In this section, we describe some of our initial research related to SCADA security and the challenges that we plan to address in the near future.

*A. Our Preliminary work*: In our preliminary work [41, 46], we have demonstrated that machine learning techniques need significant rework to perform satisfactorily in the context of anomaly detection in ICSs. The major challenge in the application of machine learning methods is obtaining real-time and unbiased datasets. Many datasets are internal and cannot be shared due to confidentiality and user privacy restrictions or may lack specific statistical characteristics. Industry settings usually avoid sharing their protected network data information because of these issues. Therefore, researchers prefer to generate datasets for training and testing purposes in the simulated or closed experimental environments that may lack comprehensiveness. Machine learning models trained with such a single dataset generally result in a semantic gap between results and their application.

As demonstrated in [41], we have obtained satisfactory results with UNSW (University of New South Wales) training and testing datasets [74] and supervised machine learning algorithms (Fig. 2) for the security of cloud platforms. However, we argue that the supervised machine learning models that perform well with a particular dataset may or may not perform suitably with completely different datasets generated with different simulation or experimental conditions. To demonstrate this, we have tested the above models with a different dataset (ISOT [68]) and observed much-degraded performance as shown in Fig. 3.

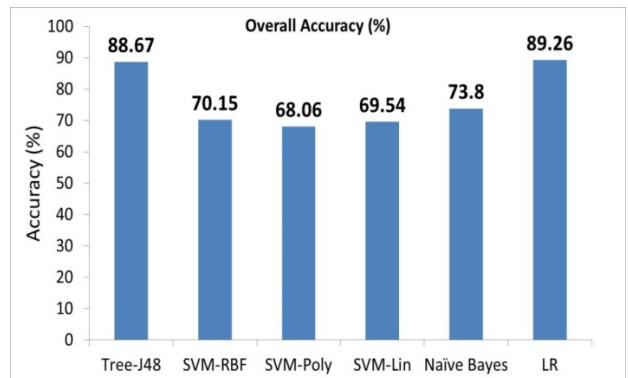
Fig. 2: Overall accuracy with UNSW Dataset.

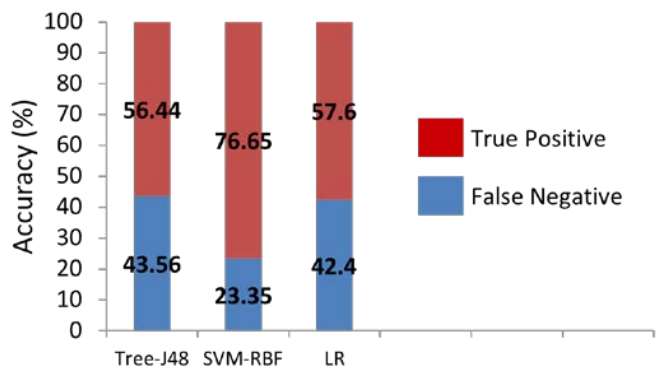
Fig. 3: True positive and true negative rates with the ISOT dataset.

The same issue applies to the application of machine learning methods to the cybersecurity of ICS as well. There is a dearth of research work, which demonstrates the effectiveness of machine learning models across multiple datasets obtained in different environments. There is a dearth of research work, which demonstrates the effectiveness of these models across multiple datasets obtained in different environments. We argue that it is necessary to test the robustness of the machine learning models, especially in diversified operating conditions, which are prevalent in cloud-based control system scenarios. Our results highlight the need for an ICS security testbed, which can be used to model real

ICSs and study the effects of attacks on them. The testbed would provide an innovative environment where researchers can explore cyber-attacks and defense mechanisms while evaluating their impact on control systems.

*B. Future Work*: As future work, we believe it is important to prepare a hybrid dataset using various major datasets, which are available online and the datasets obtained through the testbed simulation setup or at the industry campuses. We aim to extend our preliminary work in the context of the ICSs with the hybrid datasets including the one presented in [1] to test the accuracy of the machine learning techniques in the contest of ICSs.

As a part of our future research, we aim to prepare hybrid datasets using multiple publicly available datasets such as UNSW (University of New South Wales) dataset [74], ISOT dataset [68], Lawrence Berkeley National Lab (LBNL) dataset [73], Information Security Center of Excellence (ISCX) [76], next-generation IDS dataset (NGIDS-DS) [75] and other relevant datasets.

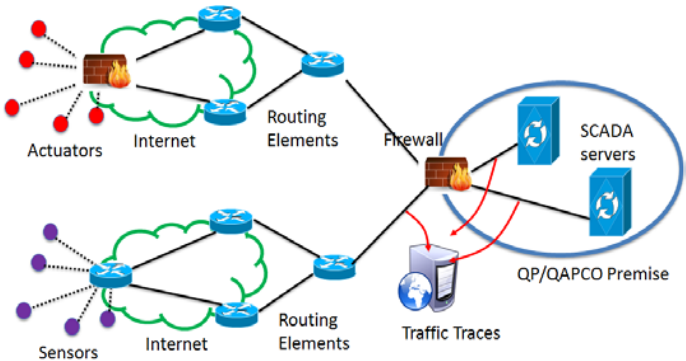

Fig. 4: Setup at an industrial premise to collect real-time traffic traces.

To achieve this, we propose to extract a common set of features from the raw TCP dumps of these datasets. This will provide the required robustness to the hybrid datasets for multi-cloud scenarios. A point to be noted here is that the datasets obtained through the lab-setups or simulated cloud environments may not represent the real-time data traffic in actual industrial control system scenarios. Therefore, in addition to preparing the hybrid dataset with the online datasets, we aim to augment the existing datasets using the real-time data traces captured at an industry campus. A possible setup at the industry premise is shown in Fig. 4, where connectivity among the remote sensors, actuators through PLC and SCADA over the Internet, using a cloud platform is shown. As shown in Fig, 4, we will collect the traffic dumps before and after the firewall as indicated in the diagram. After comparing the two dumps, malicious and normal flows can be labeled and used for the training purpose. Such dataset can be mixed with already available datasets, such as in [1] to obtain the hybrid datasets. In addition, to augment the real-time traces to prepare more robust training models, a simulation testbed for ICS may be built. A sample lab setup is shown in Fig. 5, demonstrating the interaction between PLC, SCADA, and the sensors as well as the actuators.

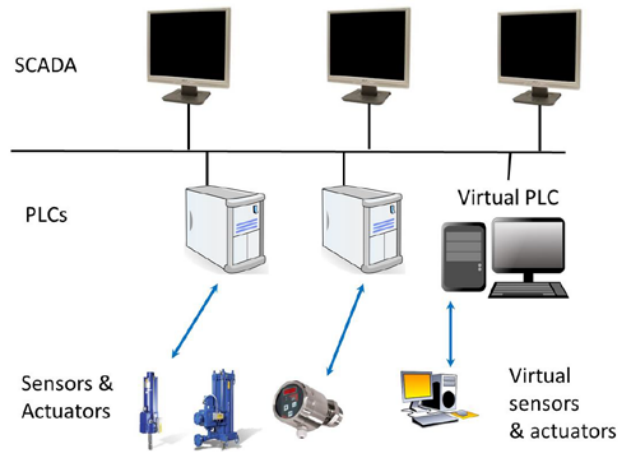

Fig. 5: Simulation testbed with the lab setup.

A sample cloud-based setup, with a possible attack scenario, is shown in Fig. 6, where a cloud is introduced for communication among various components such as remote sensors and actuators. Towards this goal, the Tennessee Eastman (TE) control process system, which is a well-known industrial process control system benchmark with well-understood dynamical behavior, can be considered.

The open-loop TE system is a complex, highly nonlinear unstable system, and hence it represents a real-world vulnerable system where cyber-attack can lead to unstable operations leading to human, environmental, and economic consequences. It should be mentioned that TE benchmark has been recently considered from cyber-security perspective [69-72] with the main focus on evaluating the behavior of the system under cyber-attack.

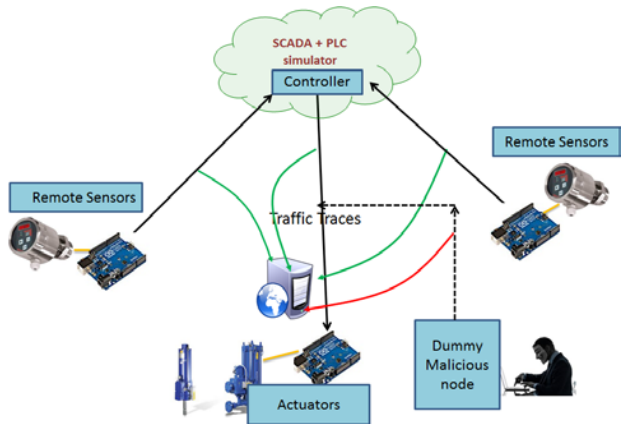

Fig. 6: Simulation testbed with cloud and infected traffic.

However, intrusion detection has not been thoroughly investigated in the literature. Consequently, it can serve as the perfect benchmark to test and validate the developed ML algorithms in future works. The effects of attacks in the process control domain can be analyzed using the simplified model of the TE process, which is mentioned in [69]. An ICS simulation testbed, we believe, can serve the following purpose:

1. Be a prototype of a common technical platform aimed at facilitating the creation of future ICS cybersecurity test centers.
2. Provide industries with a cost-effective test platform that reduces the costs of simulation and testing and delivers more significant results than a traditional testbed approach.
3. Perform cyber-attacks on a hybrid model of real ICS configuration.
4. Create a testbed that is easy to deploy, more realistic than simulation and less expensive.
5. Prepare hybrid datasets to train the machine learning models to build robust intrusion detection systems for ICSs.

## VII. Concluding Remarks

Despite the recent popularity of employing big data analytics and cloud computing for ICSs, their security is still an open issue. ICSs and eventually, the industries would benefit from the cloud platforms; however, lack of proper security in novel multi-cloud platforms may cause high costs associated with the security breaches in the real-time industry platforms. The sophistication of new malware attacking control systems, such as zero-day attacks and rootkits, has made it very difficult to prevent and detect attacks at the ICS component level. Therefore, there is a need for new schemes of intrusion detection for ICS systems at the process control level. Applicability of the machine learning techniques has proven to be very useful for this matter.

In this work, first, we took a close look at the shift of the ICS from stand-alone systems to cloud-based environments. Then we discussed the major works, from industry and academia towards the development of the secure cloud-based ICS leveraging the advancements in the field of machine learning techniques. In addition, we believe that a testbed may help to address the challenges associated with securing an industrial process, providing more insights into the knowledge about how the process is actually being managed with the help of actuators and control laws, and an understanding of the security requirements specific to process control using cloud platforms.

### Acknowledgment

This publication was made possible by the NPRP 10-0206-170360 from the Qatar National Research Fund (a member of The Qatar Foundation). The statements made herein are solely the responsibility of the author[s].

Table 2: List of Acronyms

| Acronym | Description |
| --- | --- |
| ABE | Attribute-based encryption |
| AC-i | Actuator-sensor interface |
| AES | Advanced encryption standards |
| AGA | American gas association |
| API | American petrol institute |
| APT | Advanced persistent threats |
| ASP | Application service provider |
| CIP | Common industrial protocol |
| DCS | Distributed control systems |
| DDoS | Distributed denial of service |
| DES | Data encryption standards |
| DoS | Denial of service |
| HART | Highway addressable remote transducer protocol |
| HMM | Hidden Markov Model |
| HTTP | Hypertext Transfer Protocol |
| IaaS | Infrastructure as a service |
| IAEA | International atomic energy agency |
| ICS | Industrial control system |

| IDS | Intrusion detection system |
|---|---|
| IEC | International Electrotechnical commission |
| IED | Intelligent electronic device |
| ISA | International society of automation |
| ISO | International organization for standardization |
| IT | Information Technology |
| ML | Machine learning |
| NERC | North American electric reliability corporation |
| NIDS | Network intrusion detection system |
| OCSVM | One-class support vector machine |
| PES | Power engineering society |
| PLC | Programmable logic controllers |
| PSCC | Power system communications committee |
| RTU | Remote terminal unit |
| SCADA | Supervisory control & data acquisition |
| SGX | Software guard extension |
| SOA | Service-oriented architecture |
| SVM | Support vector machine |
| TCP | Transmission control protocol |
| TE | Tennessee Eastman |
| UNSW | University of New South Wales |
| URL | Uniform Resource Locator |
| WSNM | Water supply network monitoring |